\documentclass{INTERSPEECH2023}
\usepackage{amsfonts,amssymb,pifont}
\usepackage{multirow,color,cite}

\title{A Comparative Study on Multichannel Speaker-Attributed Automatic Speech Recognition in Multi-party Meetings}
\name{Mohan Shi$^1$, Jie Zhang$^1$, Zhihao Du$^2$, Fan Yu$^2$, Qian Chen$^2$, Shiliang Zhang$^2$, Li-Rong Dai$^1$}
\address{
  $^1$NERC-SLIP, University of Science and Technology of China (USTC), China\\
  $^2$Speech Lab, Alibaba Group, China}
\email{}

\begin{document}
\maketitle
\begin{abstract}
Speaker-attributed automatic speech recognition (SA-ASR) in multi-party meeting scenarios is one of the most valuable and challenging ASR tasks. It was shown that single-channel frame-level diarization with serialized output training (\textbf{SC-FD-SOT}), single-channel word-level diarization with SOT (\textbf{SC-WD-SOT}) and joint training of single-channel target-speaker separation and ASR (\textbf{SC-TS-ASR}) can be exploited to partially solve this problem, where the latter achieves the best performance. In this paper, we propose three corresponding multichannel (MC) SA-ASR approaches, namely \textbf{MC-FD-SOT}, \textbf{MC-WD-SOT} and \textbf{MC-TS-ASR}. For different tasks/models, different 
multichannel data fusion strategies are considered, including channel-level cross-channel attention for MC-FD-SOT, frame-level cross-channel attention for MC-WD-SOT and neural beamforming for MC-TS-ASR. Experimental results on the AliMeeting corpus reveal that our proposed multichannel SA-ASR models can consistently outperform the corresponding single-channel counterparts in terms of the speaker-dependent character error rate (SD-CER).
\end{abstract}

\begin{figure*}[ht]
	\centering
	\includegraphics[width=1\linewidth]{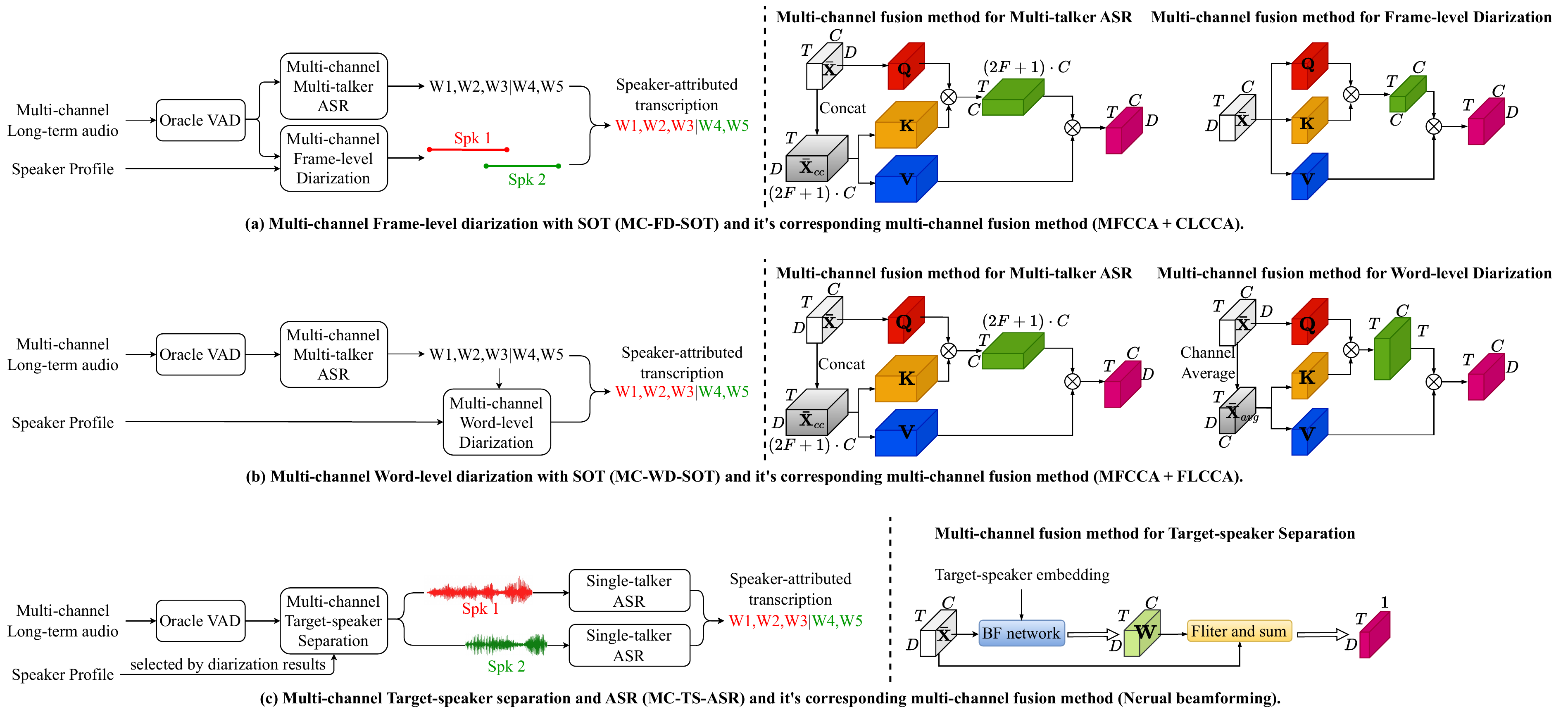}
	\vspace{-0.2cm}
	\caption{
	 The diagram of MC-FD-SOT, MC-WD-SOT and MC-TS-ASR and the corresponding simple diagram of multichannel fusion.
	}
	\label{schematic diagram}
	\vspace{-0.2cm}
\end{figure*}

\section{Introduction}
\label{sec:intro}
Multi-talker speech modeling in real meeting scenarios {(also known as the cocktail party problem) is one of} the most valuable and challenging problem in speech signal processing~\cite{fiscus2006rich,fiscus2007rich}, where speaker-attributed automatic speech recognition (SA-ASR) is expected to correctly identify ``who spoke what''~\cite{BarkerWVT18,ryant2020third,mccowan2005ami}. Therefore, SA-ASR not only needs to recognize the possible overlapping speech sounds of different speakers~\cite{yu2017recognizing,chen2017progressive,kanda2020serialized}, but also needs to identify the corresponding speaker for each recognized utterance.

{Serialized output training (SOT)~\cite{kanda2020serialized} was} proposed for multi-talker ASR in order to avoid the limit of the maximum number of speakers and duplicated hypotheses~\cite{yu2017recognizing,chang2019mimo}, which introduces a special symbol to represent the speaker change with only one output layer. To obtain the speaker attribute transcription, {three single-channel (SC) SA-ASR approaches were proposed by modular combination~\cite{YuSCSAASR}}, namely single-channel frame-level diarization with SOT (SC-FD-SOT), single-channel word-level diarization with SOT (SC-WD-SOT) and single-channel target-speaker separation and ASR (SC-TS-ASR). In SC-FD-SOT, {the frame-level speaker diarization and SOT-based ASR are combined} to obtain the speaker-attributed transcription. {{However, due to the modular independence, such an alignment strategy may suffer from erroneous timestamps which severely hinder the speaker assignment performance.}} In order to alleviate this problem, SC-WD-SOT {was} proposed, which uses an attention mechanism to diarize the hypotheses of SOT-based ASR at the token level to obtain the corresponding speaker of each character. In order to {further get rid of} the dependence on SOT-based ASR hypotheses, SC-TS-ASR was proposed {to train the joint model of target speaker separation and ASR, which achieves the best SA-ASR performance}.

Although {these approaches can work well on  real-meeting corpora, they all use single-channel audio after beamforming as the input of the model}. However, in real meeting scenarios we {usually have} multichannel signals recorded {using microphone arrays, which can be used more efficiently inside the model to learn richer temporal and spatial information.} {In this work, we therefore propose three multichannel SA-ASR approaches accordingly, called MC-FD-SOT, MC-WD-SOT and MC-TS-ASR, respectively}. 

{In literature, frame-level cross-channel attention  (FLCCA)~\cite{2021mctransformer,2021mctransducer} and channel-level cross-channel attention (CLCCA)~\cite{wang2020neural,2022Cross} were proposed in the context of ASR, speech separation and speaker diarization, where the former attends more on the time frames and the latter more on channels. Multi-frame cross-channel attention (MFCCA)~\cite{Yu2022MFCCA} reaches a tradeoff in-between by integrating cross-channel information between adjacent frames.} {As the} MFCCA-based SOT works best for the multi-talker ASR {and outperforms} the best system in M2MeT challenge~\cite{Yu2022M2MeT,Yu2022Summary,2022M2Metwin}, we adopt {the} MFCCA-based SOT to replace SOT-based ASR in SC-FD-SOT and SC-WD-SOT, namely MC-SOT. In track 1 of M2MeT, as the CLCCA {was} applied to TS-VAD to fuse multichannel speech information and {achieved} the best performance (i.e., MC-TS-VAD~\cite{2022Cross}), {we propose to combine MC-TS-VAD with MC-SOT, i.e., MC-FD-SOT.} 
{As for word-level diarization, the text and speech representations based attention is required, which means the importance of inter-frame speech context, we therefore} adopt FLCCA in word-level diarization to fuse multichannel information. Combined with multichannel word-level diarization and MC-SOT, MC-WD-SOT is then obtained. {Neural beamforming has been widely used for multichannel speech separation recently~\cite{2019fasnet,2020fasnet-tac,2021causalunet,li2022eabnet}, where a trainable neural network is used to adapt the filter weights and the single-channel output speech spectrum is then given by the filter-and-sum operation.} Considering the noise reduction capacity of neural beamforming techniques, we further propose the MC-TS-ASR method by jointly training the pre-trained neural beamforming based multichannel target speaker separation and ASR modules.

{Experimental results on a real meeting scenario corpus, AliMeeting, show that the MC-FD-SOT approach achieves an average speaker-dependent character error rate (SD-CER) of 33.5\%, which outperforms the SC-FD-SOT. The MC-WD-SOT approach achieves an average SD-CER of 30.7\%, which is better than SC-WD-SOT. The MC-TS-ASR approach obtains the best performance with an average SD-CER of 28.3\%, which is still better than SC-TS-ASR. The remainder of the paper is organized as follows. Section 2 presents the proposed MC-SA-ASR methods including MC-FD-SOT, MC-WD-SOT and MC-TS-ASR. Section 3 shows the experimental setup, followed by results in Section 4. Finally, Section 5 concludes this work.}

\section{MC-SA-ASR}
\label{MC SA-ASR}
\subsection{MFCCA-based MC-SOT}
\label{MFCCA-based MC-SOT}
The FLCCA attends more on the information between frames but simply averages the information between channels, while the CLCCA applies attention between channels but ignores the information between local frames. The MFCCA strategy can be regarded as a combination of FLCCA and CLCCA by attending both the channel-level and  frame-level information. The $i$-th head of MFCCA is calculated as
\begin{align}
\mathbf{Q}_i & =\mathbf{\bar{X}}\mathbf{W}_i^{q}+(\mathbf{B}_i^{q})^\top \in \mathbb{R}^{T \times C \times D},\\
\mathbf{K}_i & =\mathbf{\bar{X}}_{cc}\mathbf{W}_i^{k}+(\mathbf{B}_i^{k})^\top \in \mathbb{R}^{T \times (2F+1) \cdot C \times D},\\
\mathbf{V}_i & =\mathbf{\bar{X}}_{cc}\mathbf{W}_i^{v}+(\mathbf{B}_i^{v})^\top \in \mathbb{R}^{T \times (2F+1) \cdot C \times D},\\
\mathbf{H}_i &= \text{softmax} \left(\frac{  \mathbf{Q}_i\mathbf{K}_i^\top}{ \sqrt{D}} \right) \mathbf{V}_i \in \mathbb{R}^{ T \times C \times D },
\end{align}
where $T$, $C$, $D$ stand for time, channel and feature dimension, respectively, $(\cdot)^\top$ denotes transpose operation, $F$ is the number of local context frames at each time step,  $\mathbf{\bar{X}}_{cc} = [\mathbf{\bar{X}}_{cc}^{0}, \cdots, \mathbf{\bar{X}}_{cc}^T]$ with
$\mathbf{\bar{X}}_{cc}^t,\forall t\in\{0,\cdots,T\}$ being the concatenation of the context frames at time step $t$, which is constructed as $\mathbf{\bar{X}}_{cc}^t = [\mathbf{\bar{X}}^{t-F},\cdots,\mathbf{\bar{X}}^t,\cdots,\mathbf{\bar{X}}^{t+F}] \in \mathbb{R}^{ (2F+1) \cdot C\times D}$, $\mathbf{W}^{*}$ and $\mathbf{B}^{*}$ are learnable parameters.

It can be clearly seen that the MFCCA can combine both channel-level and frame-level information, {and it was shown in~\cite{Yu2022MFCCA} that the combination with the attention based encoder-decoder (AED) to configure an MFCCA-based multichannel ASR model is beneficial for performance with a small increase in the computational cost.}
 {Considering the issues of overlapping speech and unknown number of speakers in real-world meeting scenarios, the MFCCA was combined with SOT to tackle the multichannel multi-speaker ASR task}, which is called MC-SOT. It was shown that since the SOT does not require a defined number of speakers and enables to model the transcribed dependencies of different speakers and transcripts from different speakers are concatenated in line with the corresponding starting time with a separator $\langle \text{sc} \rangle$ in-between, the SOT can thus work better than permutation invariant training (PIT)~\cite{yu2017recognizing}.

\subsection{MC-FD-SOT}
\label{MC-FD-SOT}
{As the MC-TS-VAD method performs the best in the M2MeT Challenge Speaker Diarization Track~\cite{2022Cross}, which utilizes} a {CLCCA-based} TS-VAD to fuse multichannel features (e.g., the diarization error rates (DERs) are 2.26\% and 2.98\% on  AliMeeting eval and test sets, respectively). As shown in Fig.~\ref{schematic diagram}(a), here we combine MC-TS-VAD and MC-SOT by aligning their timestamps to obtain the speaker-attributed transcription. 

The detailed procedure of MC-FD-SOT is the same as {SC-FD-SOT~\cite{YuSCSAASR}. First, we estimate the utterance number $\hat{N}$ by diarization output of MC-TS-VAD and oracle sentence segmentation, and the utterance number of MC-SOT output is defined as $N$. In case} $\hat{N}$ = $N$, no further effort is required; if $\hat{N}>N$, we select the {top-$N$ longest} utterances from the output of MC-TS-VAD and discard the {rest; otherwise} we select the $\hat{N}$ utterances with the longest text length from the MC-SOT output and discard the others. Finally, we align the utterances between MC-TS-VAD and MC-SOT in a chronological order.

\subsection{MC-WD-SOT}
\label{MC-WD-SOT}
In SC-WD-SOT, three {independent} encoders are used to encode the multi-talker hypotheses, speech features and speaker embeddings. {The encoded hypotheses $H=\{\mathrm{h}_l|l=1,\dots,L\}$ and encoded} features $X=\{\mathrm{x}_t|t=1,\dots,T\}$ are input into an attention layer, {where $H$ is viewed as query, $X$ as key and value, and} the output is the word-level representation $R=\{\mathbf{r}_l|l=1,\dots,L\}$. Next, the context-independent (CI) score $S^{CI}_{l,n}$ is {calculated using} the dot product between the encoded speaker embeddings $V=\{\mathrm{v}_n|n=1,\dots,N\}$ and the aggregated representations $R$. Meanwhile, the self-attention based {network (SAN)~\cite{DBLP:journals/corr/VaswaniSPUJGKP17} is also applied to $H$ and $V$ to compute} the context-dependent (CD) score $S^{CD}_{l,n}$. Finally, the CI and CD scores are concatenated and fed to a post-processing network to predict the corresponding speaker for each character.

In MC-WD-SOT, the FLCCA is used {in prior to the} speech encoder to fuse multichannel speech information. {As shown in Fig.~\ref{schematic diagram}(b), it can better extract the frame-level speech context information while fusing the cross-channel information. FLCCA is followed by 2D-convolution to obtain the single-channel representation required by the speech encoder.} The speaker attribute transcription {can then be} obtained by merging the multichannel word-level diarization results with the MC-SOT hypotheses.

\subsection{MC-TS-ASR}
\label{MC-TS-ASR}
The multichannel target speaker separation module uses the given target speaker embedding to separate the multichannel mixed speech into {single-channel target speaker representations, and the transcription of each speaker is then obtained} by joint modeling through the ASR module, {e.g., see Fig.~\ref{schematic diagram}(c).} In general, multichannel speech separation {can be achieved by neural beamforming techniques in e.g.,}~\cite{2019fasnet,2020fasnet-tac,2021causalunet,li2022eabnet}.

{In this work, we adopt the embedding and beamforming n}etwork (EaBNet)~\cite{li2022eabnet} for multichannel target speaker separation, {{which is an effective neural beamforming approach}}. {The EaBNet mainly consists of an embedding module (EM) and beamforming module (BM), where the former follows an encoder-TCN-decoder structure}. The encoder gradually extracts features with multiple downsampling operations, {and the decoder is a mirror version of the encoder that recovers the features in the} original dimensions. Temporal convolutional networks (TCNs)~\cite{tcn} is the bottleneck connecting {the encoder and decoder, which functions as a} separator. Unlike traditional beamformers~\cite{benesty2008microphone}, the BM directly uses the mapping capability of the network to estimate the {beamformer} weights, which consists of several LSTM layers. Finally, the {desired spectrum and speech are} obtained {using the filtered spectrum and} filter-and-sum operation. In order to better extract the target speaker, feature-wise linear modulation (FILM)~\cite{perez2018film} is used to integrate the speaker embedding into the {feature} before each TCN layer.

\section{Experimental Setup}
\subsection{Dataset and evaluation metrics}
In this work, {the AliMeeting corpus~\cite{Yu2022M2MeT,Yu2022Summary} is used}, to evaluate the proposed MC-SA-ASR systems. The AliMeeting corpus contains 104.75 hours data for training (Train), 4 hours for evaluation (Eval) and 10 hours for testing (Test). Each {subset} contains several meeting sessions and each {consists of a 15 to 30-minutes} discussion by 2 to 4 participants. The dataset includes 8-channel far-field audio recorded by microphone arrays (\textit{Ali-far}) and single-channel near-field audio recorded by {participant's} headsets (\textit{Ali-near}). For MC-TS-ASR, we further simulate 50 hours mixed training data (called \textit{Train-Ali-simu}) from \textit{Train-Ali-near} to train the front-end module as well as the corresponding eval \textit{Eval-Ali-simu} and \textit{Test-Ali-simu} sets.

For SA-ASR, the speaker-dependent character error rate (SD-CER)~\cite{fu2021aishell} is used for performance evaluation. SD-CER is calculated by matching the ASR hypothesis to the corresponding speaker reference transcription, which is a rigorous evaluation metric for meeting scenarios. Besides, we  adopt the scale-invariant SNR (SI-SNR)~\cite{sisnr} to measure the speech separation performance.

\subsection{Model configuration}
{For MC-SOT, we employ a transformer model with MFCCA-based Conformer~\cite{gulati2020conformer} encoder to construct MC-SOT, which includes 11 encoder layers and 6 decoder layers with 4-head multi-head self-attention (MHSA), where the dimensions of MHSA and feed-forward network (FFN) layer are set to be 256 and 2048, respectively. The obtained CERs on Eval and Test sets of AliMeeting corpus are 17.3\% and 18.4\%, respectively. 
For MC-WD-SOT, similarly to SC-WD-SOT, a four-layer self-attention based encoder is employed to encode the recognized text with 8 attention heads and 256 hidden units in each layer. For the speech encoder, a four-layer feed-forward sequential memory network (FSMN)~\cite{zhang2015fsmn} is adopted with 256 memory blocks and a kernel size of 31. We adopt one-layer of FLCCA with 8 attention heads and 560 hidden units in prior to the speech encoder. FLCCA is followed by three-layer 2D-convolutions, where the number of kernels is \{16, 8, 1\} and the kernel size is (5, 7).
For MC-TS-ASR, to reduce the computational burden, a squeezed version of TCNs (S-TCNs) is used, where three S-TCNs are stacked and each consists of 6 squeezed temporal convolutional modules with a kernel size of 5 and dilation rate of \{1, 2, 4, 8, 16, 32\}. The feature dimension in S-TCNs is 256, and 2 uni-directional LSTM layers with 64 hidden nodes are used as the BM. We employ a Res2Net-based d-vector extraction network trained on the VoxCeleb corpus~\cite{nagrani2017voxceleb,chung2018voxceleb2} as the speaker embedding model. For fair comparison, the conformer with 8 layer encoder is used as the back-end ASR model, so that the amount of model parameters is 44M as close as that of MFCCA-based MC-SOT (45M).}

\subsection{Training details}
{T}he 80-dimensional log Mel-filter bank feature (Fbank) is {calculated as the input feature using the} ESPnet toolkit~\cite{watanabe2018espnet}. We use 4950 commonly-used Chinese characters as the modeling units. The MFCCA-based MC-SOT model is trained using \textit{Train-Ali-far} and \textit{Train-Ali-near} ({i.e., around 210 hours of speech data in total}). For MC-TS-ASR model, we first pretrain the front-end model with \textit{Train-Ali-simu} and the ASR model with \textit{Train-Ali-near}, {respectively, which are then finetuned  jointly using} \textit{Train-Ali-simu}. To maintain the front-end performance, we still adopt signal-level separation loss {for pre-training, which is linearly combined with the ASR loss, such that the total loss function can be given by}
\begin{equation}
\begin{aligned}
Loss=\lambda{L^\text{SEP}}(\hat{\boldsymbol{w}},\boldsymbol{w})+
(1-\lambda){L^\text{ASR}}(\hat{\boldsymbol{y}},\boldsymbol{y}),
\end{aligned}
\end{equation}
where ${\boldsymbol{w}}$ and ${\boldsymbol{y}}$ are {the true speech and transcription, and $\hat{\boldsymbol{w}}$ and $\hat{\boldsymbol{y}}$ are corresponding hypothesises, respectively}. $L^\text{SEP}$ and $L^\text{ASR}$ are loss {functions for separation and ASR}, respectively. {The negative SI-SNR is used as the separation loss, while joint attention-CTC loss is used for ASR, and $\lambda$ is set to be} 0.5. Finally, we use \textit{Train-Ali-far} and \textit{Train-Ali-simu} to jointly finetune the model with only {the ASR loss being taken into account. All models are trained with the} Adam optimizer~\cite{2014Adam}. When pre-training the front-end model, the {learning rate is initialized to be 1.5e-4, which will be halved if the loss does not decrease for consecutive two epochs. Besides, we use a  learning rate of 1e-3 with 10000 warmup steps to jointly fine-tune the model}.

\section{Experimental Results}
{First of all, we show the performance of the proposed MC-SA-ASR methods in Table~\ref{tab:result_pipline} in comparison with the SC-SA-ASR counterparts, which use the \textit{Ali-far-bf} data generated by the CDDMA beamformer (bf)~\cite{huang2020differential}  for training and inference.} {In order to utilize multichannel information, both MFCCA and CLCCA are used in SOT and frame-level diarization module for MC-FD-SOT, which achieves an average relative SD-CER reduction of 18.8\% in Eval and Test sets compared to SC-FD-SOT (from 41.2\% to 33.5\%). For MC-WD-SOT, MFCCA and FLCCA are used in SOT and word-level diarization module, an average relative SD-CER reduction of 16.6\% is achieved compared to SC-WD-SOT (from 36.8\% to 30.7\%).} {It is clear that the proposed MC-TS-ASR method, which calculates the filter weights of a neural beamformer and then performs filter-and-sum to obtain the fused spectrum, results in the best performance among all MC-SA-ASR systems and an average relative SD-CER reduction of 17.7\% compared to SC-TS-ASR (from 34.4\% to 28.3\%)}.

\begin{table}[!t]
\centering
\caption{The {performance comparison between MC-SA-ASR and SC-SA-ASR approaches in terms of SD-CER} (\%).}
\vspace{-0.1cm}
\setlength{\tabcolsep}{4pt}
\begin{tabular}{c|c|ccc}
\toprule
\hline
Approach & MC Fusion & Eval    & Test   & Average    \\ \hline
SC-FD-SOT~\cite{YuSCSAASR}   & CDDMA bf     & 41.0  & 41.2  & 41.2  \\
SC-WD-SOT~\cite{YuSCSAASR}   & CDDMA bf     & 36.0  & 37.1  & 36.8  \\
SC-TS-ASR~\cite{YuSCSAASR}     & CDDMA bf        &  32.5      &  35.1   & 34.4      \\ \hline
MC-FD-SOT     & MFCCA+CLCCA    &  32.8      &  33.7   & 33.5      \\
MC-WD-SOT    & MFCCA+FLCCA        & 30.7      &  30.7  & 30.7  \\
MC-TS-ASR     & Neural bf     & \textbf{30.4}   &  \textbf{27.5}   & \textbf{28.3}  \\
\hline
\bottomrule

\end{tabular}
\label{tab:result_pipline}
\end{table}


Second, we show the performance of the application of different multichannel information fusions to the proposed MC-WD-SOT method in Table~\ref{tab:result_wdsot}. To improve the performance, similarly to~\cite{YuSCSAASR} we use the combination of ASR hypotheses and real transcriptions for training. It is clear that combining the single-channel word-level diarization (SD-WD) and MC-SOT results, the performance can be largely improved compared to SC-WD-SOT. Using the FLCCA before speech encoder to fuse the multichannel information can further improve the performance, which converges to the case when an oracle separator ($\langle \text{sc} \rangle$) is additionally exploited.
\begin{table}[!tb]
\centering
\caption{The performance of the proposed MC-WD-SOT method with different multichannel fusion strategies in SD-CER (\%).}
\vspace{-0.1cm}
\setlength{\tabcolsep}{2mm}
\begin{tabular}{c|c|ccc}
\toprule
\hline
Approach   & MC Fusion   & Eval    & Test  & Average   \\ \hline
SC-WD-SOT   & CDDMA bf   & 36.0  & 37.1 &36.8\\ \hline
\multirow{3}{*}{MC-WD-SOT}  & MFCCA+SC-WD    & 30.7  &33.1   &32.4   \\
 & MFCCA+FLCCA    &30.7      &30.7     &30.7  \\
& \quad \quad + Oracle $\langle \text{sc} \rangle$    & \textbf{30.1} & \textbf{30.5}     & \textbf{30.4} \\

\hline
\bottomrule
\end{tabular}
\label{tab:result_wdsot}

\end{table}


Since {short utterances might cause interference and lead to too many insertion errors, we further show the impact of the {minimum length} of diarization utterances on the MC-TS-ASR performance} in Table~\ref{tab:result_tsasr_eval}. Compared with the original oracle speaker labels, the ones obtained from speaker diarization (MC-TS-VAD) {obtain better results in most cases}, due to the fact that there are some interfering utterances that are not recognized by the diarization model. {A further improvement can be achieved by removing the very short diarization utterances, especially on the Eval set  (e.g., from 30.4\% to} 28.2\%).

\begin{table}[!tb]
\setlength{\tabcolsep}{5pt}
\centering
\caption{{The performance (in SD-CER (\%)) of MC-TS-ASR in terms of the different minimum lengths of diarization utterances}.}
\vspace{-0.1cm}
\setlength{\tabcolsep}{2mm}
\begin{tabular}{ccccccc}
\toprule
\hline
\multicolumn{1}{c}{}    &           &   \multicolumn{4}{c}{Speaker diarization}  \\ \cline{3-7} 
\multicolumn{1}{c}{\multirow{-2}{*}{Set}}  & \multirow{-2}{*}{Oracle}   & 0s         & 0.3s          & 0.6s        & 0.9s          & 1.2s         \\ \hline
{Eval}                    & {33.2} & {30.4} & {30.1} & {28.6} & {\textbf{28.2}} & {28.5} \\ \hline
{Test}                    & {28.9} & {27.5} & {27.5} & {\textbf{27.3}} & {27.6} & {28.5} \\
\hline
\bottomrule

\end{tabular}
\label{tab:result_tsasr_eval}

\end{table}


{Finally, we compare the performance of speech separation and ASR in different stages is shown in Table~\ref{tab:result_tsasr_comp} for the proposed MC-TS-ASR method. Specifically, after pre-training we first fine-tune the whole model jointly on \textit{Train-Ali-simu}, such that the front-end separation module can adapt} to the back-end ASR module. We then fine-tune the model using real data composed of \textit{Train-Ali-far} and \textit{Train-Ali-simu}. In this step, {{we use ASR-only and joint training strategies for comparison. The difference between them is whether to freeze the front-end module during training. It can be seen that the relative SD-CER of joint training decreases by 34.2\% (from 46.2\% to 30.4\%) and 37.2\% (from 43.8\% to 27.5\%) on the Eval and Test sets, respectively, compared to the ASR-only training}}. However, joint training also leads to a decrease in the SI-SNR. In other words, joint fine-tuning enforces the front-end module to be adaptive with the ASR module at a sacrifice in the separation performance. {This implies that given a speaker embedding, the end-to-end ASR model directly recognizes the speaker's utterances rather than the separated speech in multi-talker scenarios. This also shows that a better front-end signal quality (e.g., in SI-SNR) does not necessarily mean a better back-end ASR performance in practice.}

\begin{table}[!tb]
\setlength{\tabcolsep}{5pt}
\centering
\caption{The {performance comparison of speech separation  in SI-SNR (dB) and ASR  in SD-CER (\%) in different stages for the proposed MC-TS-ASR method}.}
\vspace{-0.1cm}
\begin{tabular}{cccccc}
\toprule
\hline
\multicolumn{1}{c}{}    &     &  \multicolumn{2}{c}{SD-CER}  \    &   \multicolumn{2}{c}{SI-SNR (simu)}  \\ \cline{3-6}
\multicolumn{1}{c}{\multirow{-2}{*}{Stage}}  & \multirow{-2}{*}{Strategy}  & Eval    & Test  & Eval     & Test \\ \hline
\multirow{1}{*}{Fine-tune (simu)}  & Joint   & 73.2  &73.1    & 17.24   & 16.95  \\
 \hline
\multirow{2}{*}{Fine-tune (real)}   & ASR-only  & 46.2  & 43.8  & 17.24  & 16.95   \\
& Joint & \textbf{30.4}  & \textbf{27.5}  & -1.62  & -1.99 \\ 
\hline
\bottomrule
\end{tabular}
\label{tab:result_tsasr_comp}

\end{table}

\section{Conclusions}
\label{sec:conclusion}
In this paper, {we proposed three multichannel SA-ASR approaches in the context of multi-party meeting. {The MC-FD-SOT corresponds to SC-FD-SOT, which introduces MFCCA and CLCCA into ASR and frame-level diarization. While the MFCCA is also used in the ASR module of MC-WD-SOT, merged with FLCCA-based word-level diarization.} The MC-TS-ASR uses neural beamforming to fuse the multichannel information. It was shown that the proposed multichannel methods outperform the corresponding single-channel counterpart. The joint fine-tuning of MC-TS-ASR enables the front-end module to match the ASR module, but the separation performance drops, implying that a direct recognition of the target speaker's speech might be a more appropriate choice, particularly in multi-talker scenarios. To our knowledge, this work is the first attempt to the multichannel SA-ASR problem in real meeting scenarios. In the future, we will explore more effective multichannel multi-talker modeling methods in meeting scenarios and attempt to further take speaker localization (or direction-of-arrival estimation) into account, such that the considered SA-ASR can identify ``who spoke what from where"}.

\end{document}